\documentclass[conference]{IEEEtran} 
\ifCLASSOPTIONcompsoc
  \usepackage[nocompress]{cite}
\else
  \usepackage{cite}
\fi
\ifCLASSINFOpdf
\else
\fi
\usepackage[colorlinks=true,
            linkcolor=red,
            urlcolor=blue,
            citecolor=blue]{hyperref}
\usepackage{amsmath}
\usepackage{textcomp}
\usepackage{siunitx}
\usepackage{mathtools}
\usepackage{relsize}
\usepackage{graphicx}
\usepackage{algorithm}
\usepackage{filecontents,lipsum}
\usepackage{cite}
\usepackage{comment}
\usepackage{multirow}
\usepackage{amsfonts}
\usepackage[noend]{algpseudocode}
\usepackage[utf8]{inputenc}
\usepackage{graphicx}
\usepackage{caption}
\usepackage{latexsym}
\usepackage{etoolbox}
\usepackage{subcaption, float}
\makeatletter
\usepackage[inline]{enumitem}
\def\BState{\State\hskip-\ALG@thistlm}
\makeatother

\AfterEndEnvironment{table}{\vskip-1ex}

\makeatletter
\newcommand{\mypm}{\mathbin{\mathpalette\@mypm\relax}}
\newcommand{\@mypm}[2]{\ooalign{%
  \raisebox{.1\height}{$#1+$}\cr
  \smash{\raisebox{-.6\height}{$#1-$}}\cr}}
\makeatother
\begin{document}

\title{Hierarchical Deep Convolutional Neural Networks for Multi-category Diagnosis of Gastrointestinal Disorders on Histopathological Images}

%
%

\markboth{IEEE Transactions on Medical Imaging}%
{Shell \MakeLowercase{\textit{et al.}}: Bare Demo of IEEEtran.cls for IEEE Journals}
%




\author{
\IEEEauthorblockN{
Rasoul Sali$^{1}$,
Sodiq Adewole$^{1}$, 
Lubaina Ehsan$^{2}$,
Lee A. Denson$^{4}$,
Paul Kelly$^{5,6}$,
Beatrice C. Amadi$^{6}$,\\
Lori Holtz$^{7}$,
Syed Asad Ali$^{8}$,
Sean R. Moore$^{2}$,
Sana Syed$^{2,*}$, and
Donald E. Brown$^{1,3,*}$}
\\

\IEEEauthorblockA{$^{1}$ Department of Systems and Information Engineering, University of Virginia,
Charlottesville, VA, USA}

\IEEEauthorblockA{$^{2}$ Department of Pediatrics, School of Medicine, University of Virginia, Charlottesville, VA, USA}

\IEEEauthorblockA{$^{3}$ School of Data Science, University of Virginia, Charlottesville, VA, USA}

\IEEEauthorblockA{$^{4}$ Division of Gastroenterology, Hepatology, and Nutrition, Cincinnati Children's Hospital Medical Center, Cincinnati, OH, USA}

\IEEEauthorblockA{$^{5}$ Blizard Institute, Barts and The London School of Medicine, Queen Mary University of London, London, United Kingdom}

\IEEEauthorblockA{$^{6}$ Tropical  Gastroenterology  and  Nutrition  group,  University  of  Zambia, School of Medicine, Lusaka, Zambia}

\IEEEauthorblockA{$^{7}$ Department of Pediatrics, School of Medicine, Washington University, Saint Louis, MO, USA}

\IEEEauthorblockA{$^{8}$ Department of Pediatrics and Child Health, Aga Khan University, Karachi, Pakistan}
\\
$^*$co-corresponding authors:\{\href{mailto:sana.syed@virginia.edu}{sana.syed},
\href{mailto:brown@virginia.edu}{brown}\}@virginia.edu\vspace{-15pt}}

\maketitle

\begin{abstract}~Deep convolutional neural networks~(CNNs) have been successful for a wide range of computer vision tasks, including image classification. A specific area of the application lies in digital pathology for pattern recognition in the tissue-based diagnosis of gastrointestinal~(GI) diseases. This domain can utilize CNNs to translate histopathological images into precise diagnostics. This is challenging since these complex biopsies are heterogeneous and require multiple levels of assessment. This is mainly due to structural similarities in different parts of the GI tract and shared features among different gut diseases. Addressing this problem with a flat model that assumes all classes~(parts of the gut and their diseases) are equally difficult to distinguish leads to an inadequate assessment of each class. Since the hierarchical model restricts classification error to each sub-class, it leads to a more informative model than a flat model. In this paper, we propose to apply the hierarchical classification of biopsy images from different parts of the GI tract and the receptive diseases within each. We embedded a class hierarchy into the plain VGGNet to take advantage of its layers’ hierarchical structure. The proposed model was evaluated using an independent set of image patches from \textbf{$373$} whole slide images. The results indicate that the hierarchical model can achieve better results than the flat model for multi-category diagnosis of GI disorders using histopathological images.\vspace{5pt}
\end{abstract}


\section{Introduction}\label{sec:Introduction}
\vspace{-2pt}
Gastrointestinal (GI) diseases are ailments linked to the digestive system, including the esophagus, stomach, and the intestines. GI diseases account for substantial morbidity, mortality, and financial burden by affecting the GI tract and impacting digestion and overall health. The National Institute of Health reports that between 60 and 70 million Americans are affected by GI diseases each year\cite{peery2012burden}.

A common approach to GI disease diagnosis lies in digital pathology for pattern recognition. However, a significant challenge of interpreting clinical biopsy images to diagnose disease is the often striking overlap in histopathology images between distinct but related conditions. There is a critical clinical need to develop new methods to allow clinicians to translate heterogeneous biomedical images into precise diagnostics~\cite{bejnordi2017diagnostic}.

Convolutional Neural Networks (CNNs) have shown superior performance for the automated extraction of quantitative morphologic phenotypes from GI biopsy images and associated diseases diagnosis \cite{sali2019celiacnet,kowsari2019diagnosis}. Despite this, as the number of disease classes associated with different parts of the gut becomes larger, one of the problems that may arise is that visual separability between classes becomes more challenging. This is mainly due to structural similarities in different gut parts and the shared features among gut diseases. Furthermore, for multi-class classification problems, some classes become harder to distinguish than others and require dedicated classifiers~\cite{yan2015hd}. Regular flat models cannot address this issue because they assume that all classes are equally difficult to distinguish~\cite{zhu2017b}. Hierarchical relationships are often identified that exist between classes, which can be used to deploy a hierarchical classification model. This model can be more informative since the classification error is restricted to subcategories compared to treating all classes as arranged in a flat structure. 

In CNNs architecture, lower layers capture low-level features while higher layers are likely to extract more abstract features~\cite{zeiler2014visualizing}. This CNN property can be combined with the hierarchical structure of classes to enforce the network to learn different levels of class hierarchy in different layers. In this way, coarse categories that are easier to classify are represented in lower~(shallow) layers while higher~(deeper) layers output fine subcategories simultaneously~\cite{yan2015hd,zhu2017b}. In this paper, we propose a hierarchical deep convolutional neural network to take advantage of GI diseases’ hierarchical structure for classification. 

This paper is organized as follows: Section~\ref{sec:diseases} provides an introduction to the diseases studied in this paper. In section ~\ref{sec:RelatedWork} some related researches are reviewed. The methodology is explained in section~\ref{sec:Method}. The data used in this study, data preparation steps and empirical results are elaborated in section~\ref{sec:EmpiricalResults}. Finally, section~\ref{sec:Conclusion} concludes the paper along with outlining future directions.

\section{Gastrointestinal disorders}\label{sec:diseases}
GI disorders refer to any abnormal condition or disease that occurs within the GI tract. While there is a wide variety of disorders associated with different parts of the GI tract, this paper focuses on certain disorders involving only the duodenum, esophagus, and ileum. In this section, we give an introduction to each of the considered disorders.
\subsection{Duodenum}
\subsubsection{Celiac Disease (CD)} 
It is an inability to normally process dietary gluten (present in foods such as wheat, rye, and barley) and is present in $1$\% of the US population. Gluten exposure triggers an inflammatory cascade, which leads to a compromised intestinal barrier function. Gluten consumption by people with CD can cause diarrhea, abdominal pain, bloating, and weight loss. If unrecognized, it can lead to anemia, decreased bone density, and, in longstanding cases, intestinal cancer~\cite{parzanese2017celiac}.

\subsubsection{Environmental Enteropathy (EE)}
It is an acquired small intestinal condition resulting from the continuous burden of immune stimulation by fecal-oral exposure to enteropathogens leading to a persistent acute phase response and chronic inflammation~\cite{campbell2003growth,solomons2003environmental}. EE can be characterized histologically by villus shortening, crypt hyperplasia, and a resultant decrease in the surface area of mature absorptive intestinal epithelial cells, leading to a markedly reduced nutrient absorption, under-nutrition, and stunting~\cite{syed2016environmental}.

\subsection{Esophagus}
\subsubsection{Eosinophilic Esophagitis (EoE)}
It is a chronic, allergic inflammatory disease of the esophagus. It occurs when eosinophils, a normal type of white blood cells present in the digestive tract, build up in the lining of the esophagus. EoE is characterized by symptoms of esophageal dysfunction and eosinophilic infiltration of the esophageal mucosa in the absence of secondary causes of eosinophilia~\cite{dellon2018epidemiology}.

\subsection{Ileum}
\subsubsection{ Crohn’s Disease}
It is an inflammatory bowel disease that causes patchy disease constituting of chronic inflammation, ulcers, and mucosal damage anywhere in the GI tract, although the most common being the terminal ileum and colon. The interaction of genetic susceptibility, environmental factors,  and intestinal microflora is believed to be the major cause of Crohn’s disease. This interaction results in abnormal mucosal immune response, which compromises epithelial barrier function~\cite{torres2017crohn}.

\section{Related Work}\label{sec:RelatedWork}
Hierarchical CNN has demonstrated improved performance in image classification compared to flat CNN models across multiple domains~\cite{yan2019hierarchical, seo2019hierarchical, milletari2016v, ranjan2018hierarchical}. These models exploit the hierarchical structure of object categories~\cite{tousch2012semantic} to decompose the classification tasks into multiple steps. Hierarchical Deep CNNs (HD-CNN)  proposed by Yan et al.~\cite{yan2015hd} embeds CNN into a categorical hierarchy by separating easy classes using a coarse category classifier and difficult categories using a fine category classifier. This model can be implemented without increasing the complexity of the training process; however, it requires multi-step training of each CNN. Zhu and Bain proposed a branched variant (B-CNN) of the hierarchical deep CNN~\cite{zhu2017b}. Since shallow layers of a CNN capture low-level features while deeper levels capture high-level features, B-CNN outputs multiple predictions ordered from coarse to fine along concatenated convolutional layers corresponding to the hierarchical structure of the target classes. The model branch training strategy adjusts parameters on the output layers, forcing the input to learn successively coarse to fine concepts along with the layer blocks. Hierarchical architecture has been applied to both image~\cite{seo2019hierarchical} and video classification~\cite{fan2004classview} tasks with superior performance compared with conventional flat CNN models. While deep learning has seen a significant application in medical image classification tasks~\cite{milletari2016v}, hierarchical models remain a relatively less explored area in literature. 

Ranjan et al. reported a CNN-based hierarchical model in medical image classification on histopathological images.~\cite{ranjan2018hierarchical} with superior performance than flat CNN. They proposed to classify cancer and its states (in situ, invasive, or normal) using multiple CNNs organized hierarchically. With one CNN in each of the two-level classification tasks, the first level CNN, a pre-trained AlexNet, is trained to discriminate the normal class from the rest of the classes’ images. The second level of the hierarchy is a tree of CNN-based binary classifiers using majority voting to discriminate the other three classes; in situ, invasive and benign cells.

Krauß et al. applied a hierarchical CNN on cytopathology to classify cellular images as healthy or cancer‐affected cells~\cite{krauss2018hierarchical}. To the best of our knowledge, there are no previously published studies that have applied hierarchical deep convolutional neural networks to gastrointestinal disease classification using histopathological images.

\section{Methodology}\label{sec:Method} 
\subsection{Base Model}
There are many different architectures of CNNs in the literature with associated advantages and drawbacks. In this paper, we used VGGNet~\cite{simonyan2014very}~(proposed by Visual Geometry Group in the University of Oxford) as a base model that has shown excellent performance in image classification problems, including medical image analysis~\cite{rakhlin2018deep,gomez2019automatic,ko2019deep}. VGGNet obtained the state-of-the-art results in the ILSVRC’14 competition with $7.3\%$ error rate, which was among the top $5$ errors and was a significant improvement over ZFNet~\cite{zeiler2014visualizing}, the winner of ILSVRC’13. Two main intertwined characteristics of VGGNet were the increased depth of the network and applying smaller filters. It uses $3\times3$ sized filters and $2\times2$ sized pooling from the beginning to the end of the network. Since smaller filters have few parameters, it is possible to increase the depth by stacking more of them with the same effective receptive fields when using larger filters. For instance, effective receptive fields of three stacked $3\times3$ filters with stride $1$ are the same as a $7\times7$ filter.
VGG16 and VGG19 have been released as two variants of VGGNet. VGG16 has $16$ trainable layers, including $13$ convolutional layers organized in $5$ blocks followed by $3$ fully-connected layers. The final layer is a softmax layer that outputs class probabilities. In this paper, VGG16 was applied and was trained from scratch on biopsy patches. The network is shown at the top in Figure \ref{fig:model} is a plain VGG16.

\subsection{Hierarchical Convolutional Neural Network}
\begin{figure*}[!htb]
    \centering
    \includegraphics[width=0.97\textwidth]{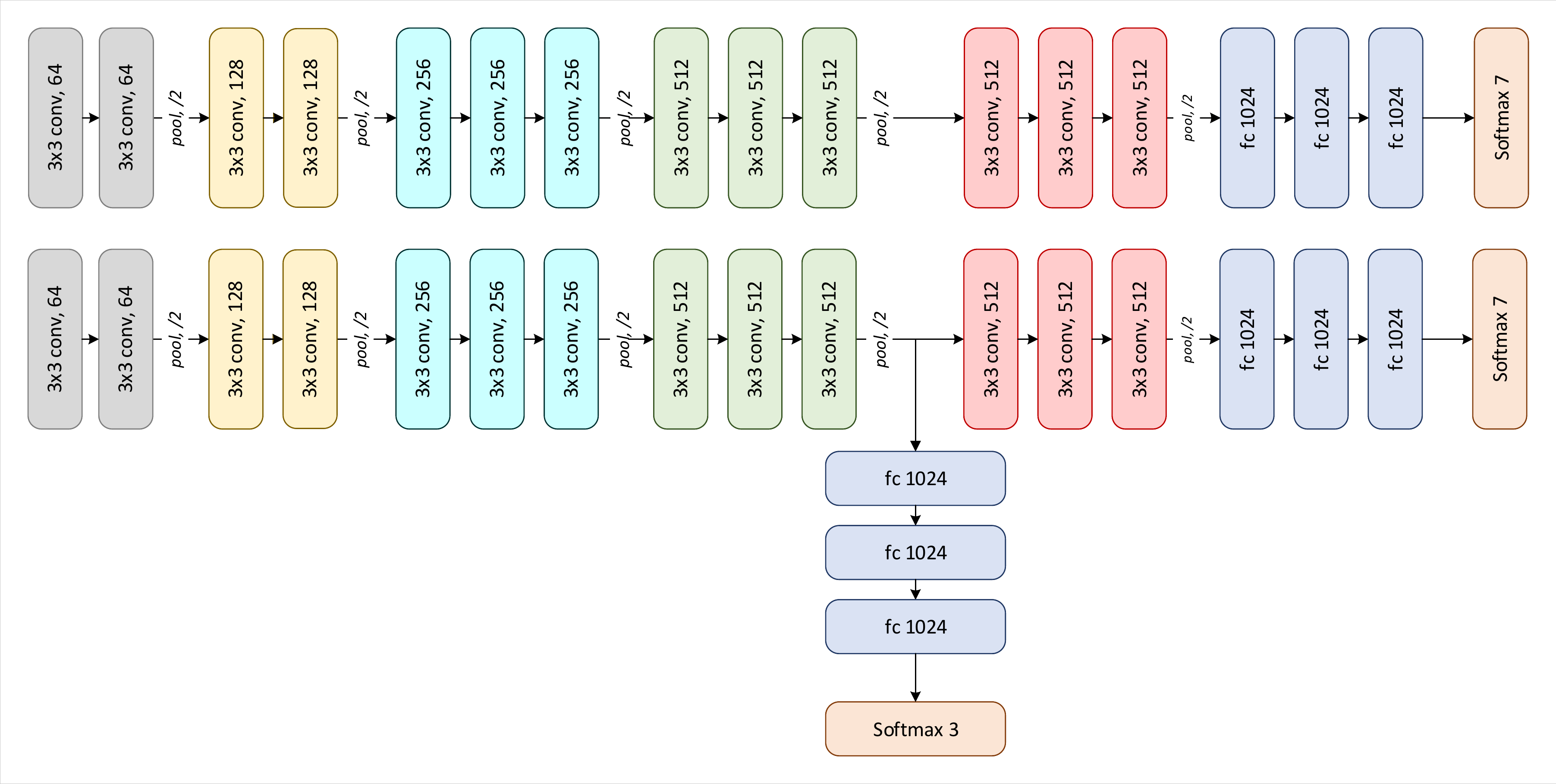}
    \caption{ Top: VGGNet architecture, Bottom: H-VGGNet architecture} \label{fig:model}
\vspace{-5pt}
\end{figure*}

\begin{figure}
    \centering
    \includegraphics[width=0.98\columnwidth]{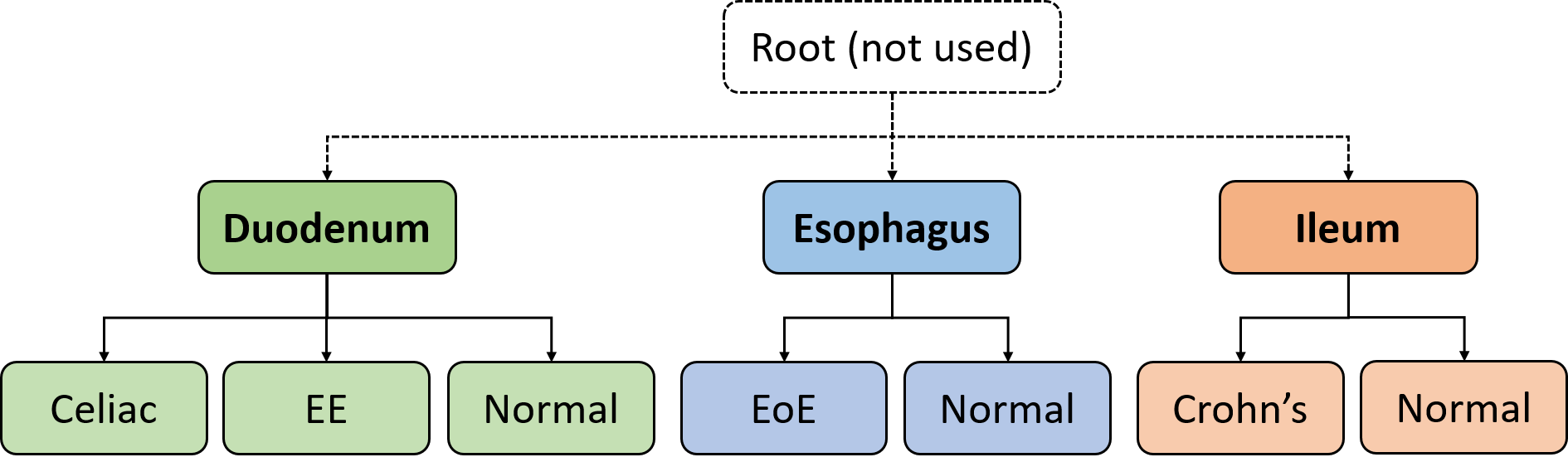}
    \caption{Hierarchy of classes} \label{fig:LabelsTree}
\end{figure}

To propose a hierarchical convolutional neural network, the architecture of Branch Convolutional Neural Network (B-CNN)~\cite{zhu2017b} proposed by Zhu and Bain was applied to embed different levels of class hierarchical on VGG16 to propose H-VGGNet. There were seven classes of GI disorders: Duodenum-Celiac, Duodenum-EE, Duodenum-Normal, Esophagus-EoE, Esophagus-Normal, Ileum-Crohn’s and Ileum-Normal, as fine classes and each class belonged to a coarse category: Duodenum, Esophagus or Ileum (see the hierarchy of classes in Figure \ref{fig:LabelsTree}). Since there are two class levels in this hierarchy, in addition to output layer, which will output fine classes, one branch was added to VGG16 to output the coarse categories. The network is shown at the bottom in Figure \ref{fig:model} is H-VGGNet. Although new branches can consist of both convolutional and fully-connected layers, in this paper, the new added branch was composed of only fully connected layers. This model for each input image computed both coarse and fine level predictions. The rectified linear unit~(ReLU)~\cite{krizhevsky2012imagenet} was employed as the activation function. To reduce over-fitting, dropout regularization~\cite{srivastava2014dropout} was used after ReLU of each fully-connected layer with $p=0.5$. Also, Batch Normalization~\cite{ioffe2015batch} was applied after ReLU of every trainable layer.

The loss function for such a model is weighted summation of coarse and fine prediction losses (see equation \ref{eq:loss}).

\begin{equation}\label{eq:loss}
\mathcal{L}_{i} = -\sum_{k=1}^{K}{w_{k}\log(\frac{e^{s_{y_{ip}}^{k}}}{\sum_{j}{e^{s_{y_{ij}}^{k}}}})}
\end{equation}

Where $K$ is number of levels in hierarchy of classes, $w_{k}$ is weight of level $k$th in class hierarchy and term $-\log(.)$ is cross-entropy loss function for the $i$th instance in $k$th level of class hierarchy. $s_{y_{ip}}^{k}$ is the element in class score of instance $i$th in $k$th level of class hierarchy corresponding to positive element in target label $y_{i}$ and $s_{y_{ij}}^{k}$ is the $j$th element in class score of instance $i$th in $k$th level of class hierarchy.

\section{Experimental Setup}\label{sec:EmpiricalResults}
This section is devoted to presenting the experimental setting, including data description, data pre-processing steps, training details, and the evaluation criterion.
\subsection{Data}\label{sec:Data_Source}
$1,150$ Hematoxylin and Eosin~(H\&E) stained whole slide images (WSI) from $441$ patients were obtained for this study. Some patients had $2-3$ biopsies for each diagnosis. All biopsies obtained from the University of Virginia (UVA), VA, USA were retrospectively retrieved archival samples while the other biopsies were obtained as part of prospective cohorts studying growth faltering among children (except Crohn’s Disease biopsies). Images were obtained based on the gut disease state: 1) Celiac Disease in the Duodenum: UVA ($n=239$), Cincinnati Children’s Medical Center (CCHMC), OH, USA and Washington University (WashU), MO, USA  ($n=43$ and $n=54$, respectively); 2) Environmental Enteropathy in Duodenum: Aga Khan University, Karachi, Pakistan ($n=34$) and Zambia School of Medicine, Lusaka, Zambia  ($n=19$); 3) histologically normal duodenum: UVA ($n=174$), CCHMC ($n=36$), WashU ($n=11$); 4) EoE in Esophagus: UVA ($n=349$); 5) histologically normal esophagus: UVA ($n=155$); 6) Crohn’s disease in Ileum: CCHMC as part of RISK study sub-cohort ($n=20$); and, 7) histologically normal ileum: UVA ($n=16$).\newline
We split our data into training, development, and test sets using $50:20:30$ ratio. Since the model must generalize to other unseen patients data, we performed our split to ensure no overlap between the training, development, and test set for a particular patient.

\begin{table*}[]
\centering
\caption{Distribution of training, development, and test set data among different classes}
\vspace{-3pt}
\label{tab:dataset}
\begin{tabular}{llcccccc}
\hline
\multirow{2}{*}{\textbf{Coarse category~~~}} & \multirow{2}{*}{\textbf{Fine class~~~}} & \multicolumn{2}{c}{\textbf{Train}} & \multicolumn{2}{c}{\textbf{Development}} & \multicolumn{2}{c}{\textbf{Test}} \\ \cline{3-8} 
 &  & \textbf{\begin{tabular}[c]{@{}c@{}}~Number of~ \\ WSIs\end{tabular}} & \textbf{\begin{tabular}[c]{@{}c@{}}~~Number of~~ \\ patches\end{tabular}} & \textbf{\begin{tabular}[c]{@{}c@{}}~~Number of~~ \\ WSIs\end{tabular}} & \textbf{\begin{tabular}[c]{@{}c@{}}~Number of~ \\ patches\end{tabular}} & \textbf{\begin{tabular}[c]{@{}c@{}}~~Number of~~ \\ WSIs\end{tabular}} & \textbf{\begin{tabular}[c]{@{}c@{}}~~Number of~~ \\ patches\end{tabular}} \\ \hline
\multirow{3}{*}{Duodenum} & Celiac & $170$ & $8521$ & $62$ & $1548$ & $104$ & $2738$ \\ \cline{2-8} 
 & EE & $28$ & $8138$ & $11$ & $1058$ & $14$ & $1343$ \\ \cline{2-8} 
 & Normal & $120$ & $8290$ & $36$ & $810$ & $65$ & $1257$ \\ \hline
\multirow{2}{*}{Esophagus} & EoE & $140$ & $12342$ & $75$ & $5389$ & $134$ & $10907$ \\ \cline{2-8} 
 & Normal & $80$ & $9115$ & $30$ & $3170$ & $45$ & $5202$ \\ \hline
\multirow{2}{*}{Ileum} & Crohns & $11$ & $5006$ & $3$ & $1300$ & $6$ & $1953$ \\ \cline{2-8} 
 & Normal & 8 & $3906$ & $3$ & $1312$ & $5$ & $2182$ \\ \hline
\end{tabular}
\end{table*}

\subsection{Data Pre-processing}\label{sec:Pre-Processing}
\subsubsection{Image Patching}
A sliding window method was applied to each high-resolution WSI to generate patches of size $1000\times1000$ pixels. Since some classes had more whole-slide images than others,  we generated patches with different overlapping areas for each class. Patches were resized to $224\times224$ pixels to reduce the computational cost. After generating tissue tiles from each WSI, tiles’ labels were assumed to be the same with its associated WSI.
\vspace{5pt}
\subsubsection{Patch Clustering}\label{subsec:Clustering}
\vspace{-2pt}
In this work, a two-step clustering process was applied to filter useless patches which had mostly been created from the WSIs’ background. A convolutional auto-encoder (CAE) was employed to map each patch into an embedding space through the first step. In the second step, k-means clustering was applied to cluster embedded features into two clusters: useful and useless. Table~\ref{tab:dataset} summarizes distribution of WSIs and patches (after cleaning) in each class.  
\vspace{5pt}
\subsubsection{Stain Color Normalization}\label{subsec:StainNormalization}~Histological images have substantial color variation that adds bias while training the model. This issue arises due to a wide variety of factors, such as differences in raw materials and manufacturing techniques of stain vendors, staining protocols of labs, and color
responses of digital scanners~\cite{vahadane2016structure}. Unwanted color variations should be addressed and resolved as an essential pre-processing step before any analyses to prevent any bias arising from this issue. 

\begin{figure}
    \centering
    \includegraphics[width=0.98\columnwidth]{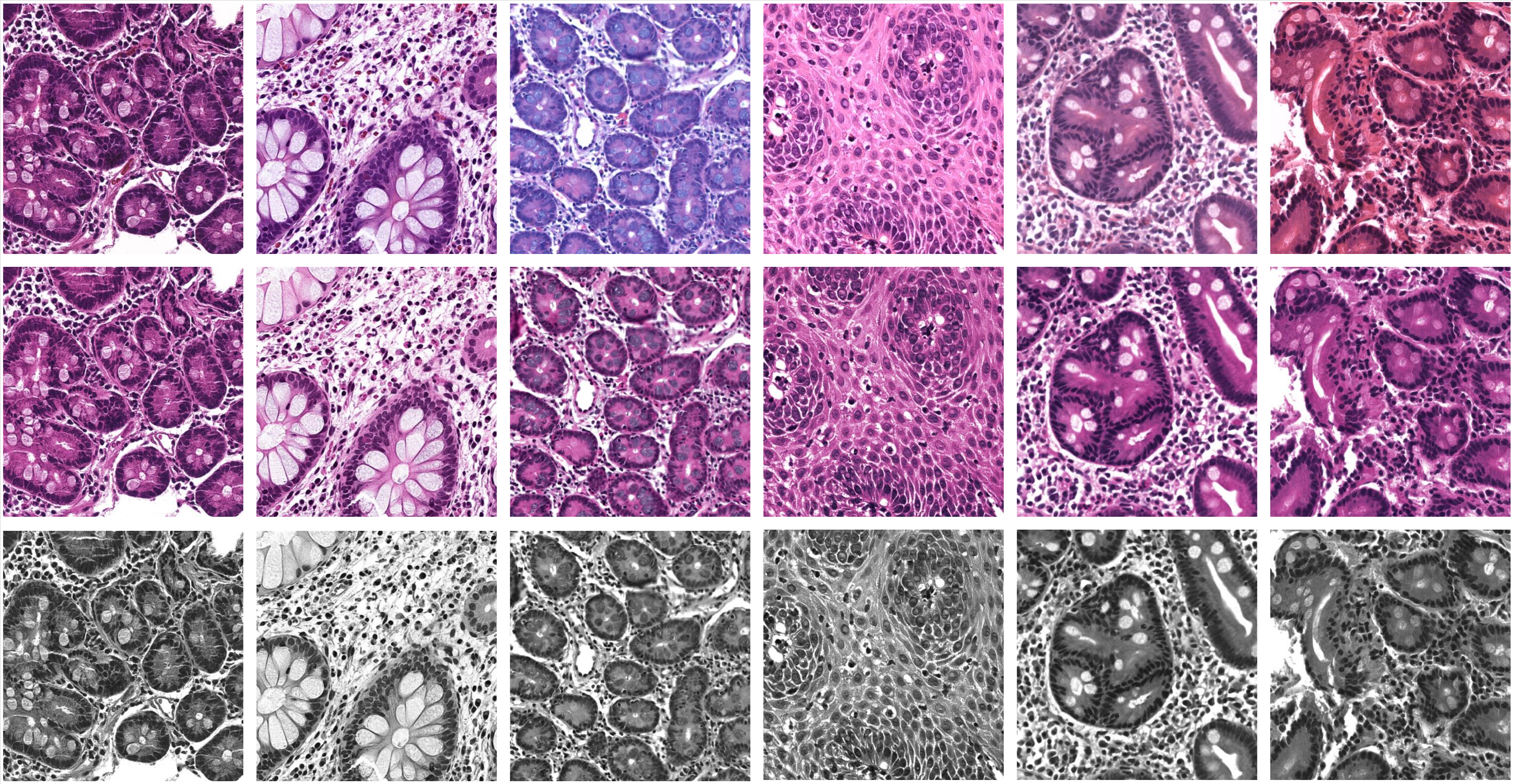}
    \caption{Color normalization artifacts. The first row represents the original images, the second row is color normalized images using the method proposed by Vahadane et al.~\cite{vahadane2016structure} and their associated gray-scale images are in the third row.} \label{fig:gray-scale}
\end{figure}

Various solutions such as color balancing~\cite{kowsari2019diagnosis}, gray-scale, and stain normalization~\cite{sali2019celiacnet} have been proposed in the published literature to address the color variation issue. In this study, we used gray-scale images. However, before converting the RGB patches to gray-scale, the stain normalization approach proposed by Vahadane et al.~\cite{vahadane2016structure} was applied to make sure that the effect of variation of color variation is significantly reduced. Figure~\ref{fig:gray-scale} shows an example of the result of applying this process on representative biopsy patches.

\subsection{Training Details}
We conducted extensive experiments to compare the performance of the hierarchical model with a flat model.
Both the base model and the hierarchical model were trained and tested ten times. Each time both the models were trained in $20$ epochs. Optimization was performed using RMSprop~\cite{tieleman2012lecture} optimization with no momentum, The  initial  value  of  the  learning  rate  is  considered  as \num{1e-3}, it changed to \num{5e-4} after the 10th epoch and to \num{1e-4} after the 15th epoch. Different loss weights are applied to each level of the hierarchy to reflect the differences in each level of classes’ importance. Since the low-level feature extraction is more important in initial epochs, more weights are assigned to it. As the model’s training progresses, the weight of the coarse categories level decreases, and the weight of fine classes increases. The changes in loss weights follow $[0.98, 0.02]$ in the first epoch, $[0.30, 0.70]$ in the 5th epoch, $[0.10, 0.90]$ in the 10th epoch, $[0.00, 1.00]$ in the 15th epoch. This change in weights causes the algorithm to focus first on the optimization of the coarse category, and as the learning process progresses, this focus shifts to the fine level.

\subsection{Evaluation Metrics}
In order to assess the performance of models, accuracy, area under the ROC curve (AUC), Precision, Recall, and F1 score have been considered.
\section{Results}
Table~\ref{tab:metrics} presents the performance comparison between two models in terms of the accuracy, AUC, Precision, Recall, and F1 score on the test set with $95\%$ confidence intervals. As shown, the hierarchical model’s performance for many classes was better than the flat model in terms of the mean of the criterion mentioned above. Also Table~\ref{tab:conf-matrix} presents the normalized confusion matrix of two models. The confusion between different coarse categories in the hierarchical model was less than the flat model.

\begin{table*}[]
\centering
\caption{Comparison of models’ performance}
\vspace{-3pt}
\label{tab:metrics}
\resizebox{0.97\textwidth}{!}{%
\begin{tabular}{llccccccc}
\hline
\multirow{3}{*}{\textbf{Metric}} & \multirow{3}{*}{\textbf{Model}} & \multicolumn{7}{c}{\textbf{Class}} \\ \cline{3-9} 
 &  & \multicolumn{3}{c}{\textbf{Duodenum}} & \multicolumn{2}{c}{\textbf{Esophagus}} & \multicolumn{2}{c}{\textbf{Ileum}} \\ \cline{3-9} 
 &  & \textbf{Celiac} & \textbf{EE} & \textbf{Normal} & \textbf{EoE} & \textbf{Normal} & \textbf{Crohn's} & \textbf{Normal} \\ \hline
\multirow{2}{*}{Accuracy} & VGGNet &
$0.941~\mypm 0.011$&
$0.986~\mypm 0.008$&
$0.959~\mypm 0.009$&
$0.917~\mypm 0.015$&
$0.928~\mypm 0.013$&
$0.960~\mypm 0.011$& 
$0.969~\mypm 0.012$ \\
 & H-VGGNet & 
 $\mathbf{0.953}\mypm 0.007$&
 $\mathbf{0.987}\mypm 0.003$&
 $\mathbf{0.963}\mypm 0.008$&
 $\mathbf{0.937}\mypm 0.012$&
 $\mathbf{0.945}\mypm 0.013$&
 $\mathbf{0.973}\mypm 0.009$& 
 $\mathbf{0.971}\mypm 0.010$ \\ \hline
\multirow{2}{*}{AUC} & VGGNet & 
$0.842~\mypm 0.039$&
$0.981~\mypm 0.011$&
$0.843~\mypm 0.032$&
$0.906~\mypm 0.016$& 
$0.923~\mypm 0.008$& 
$0.947~\mypm 0.024$& 
$0.863~\mypm 0.053$ \\ 
 & H-VGGNet & 
 $\mathbf{0.870}\mypm 0.019$ &
 $\mathbf{0.992}\mypm 0.003$ &
 $\mathbf{0.872}\mypm 0.025$ &
 $\mathbf{0.938}\mypm 0.005$ &
 $\mathbf{0.931}\mypm 0.008$ &
 $\mathbf{0.967}\mypm 0.008$ &
 $\mathbf{0.874}\mypm 0.022$ \\ \hline
\multirow{2}{*}{Precision} & VGGNet & $0.763\mypm 0.052$ &
$0.837~\mypm 0.099$ &
$0.625~\mypm 0.070$ &
$\mathbf{0.965}\mypm 0.007$ &
$0.775~\mypm 0.046$ &
$0.710~\mypm 0.089$ &
$\mathbf{0.961}\mypm 0.024$ \\
 & H-VGGNet & 
 $\mathbf{0.850}\mypm 0.020$ &
 $\mathbf{0.915}\mypm 0.018$ &
 $\mathbf{0.682}\mypm 0.053$ &
 $0.942~\mypm 0.013$ &
 $\mathbf{0.856}\mypm 0.025$ &
 $\mathbf{0.798}\mypm 0.038$ &
 $0.937~\mypm 0.013$ \\ \hline
\multirow{2}{*}{Recall} & VGGNet & 
$0.712~\mypm 0.083$ &
$0.975~\mypm 0.025$ &
$0.711~\mypm 0.071$ & 
$0.834~\mypm 0.035$ &
$\mathbf{0.915}\mypm 0.013$ &
$0.928~\mypm 0.049$ &
$0.729~\mypm 0.107$ \\
 & H-VGGNet & 
 $\mathbf{0.756}\mypm 0.041$ &
 $\mathbf{0.989}\mypm 0.005$ &
 $\mathbf{0.764}\mypm 0.055$ &
 $\mathbf{0.920}\mypm 0.017$ &
 $0.901~\mypm 0.023$ &
 $\mathbf{0.955}\mypm 0.019$ &
 $\mathbf{0.752}\mypm 0.045$ \\ \hline
\multirow{2}{*}{F1 score} & VGGNet & $0.728\mypm 0.039$ &
$0.893~\mypm 0.058$ &
$0.653~\mypm 0.023$ &
$0.894~\mypm 0.019$ &
$0.838~\mypm 0.025$ &
$0.797~\mypm 0.056$ &
$0.820~\mypm 0.033$ \\
 & H-VGGNet & 
 $\mathbf{0.799}\mypm 0.020$ &
 $\mathbf{0.950}\mypm 0.008$ &
 $\mathbf{0.714}\mypm 0.015$ &
 $\mathbf{0.930}\mypm 0.006$ &
 $\mathbf{0.877}\mypm 0.006$ &
 $\mathbf{0.868}\mypm 0.021$ &
 $\mathbf{0.833}\mypm 0.027$ \\ \hline
\end{tabular}%
}
\end{table*}

\begin{table*}[]
\centering
\caption{Normalized confusion matrix of flat and hierarchical model}
\vspace{-3pt}
\label{tab:conf-matrix}
\resizebox{0.97\textwidth}{!}{%
\begin{tabular}{llccccccc}
\hline
\multirow{3}{*}{\textbf{True Label}} & \multirow{3}{*}{\textbf{Model}} & \multicolumn{7}{c}{\textbf{Predicted Label}} \\ \cline{3-9} 
 &  & \multicolumn{3}{c}{\textbf{Duodenum}} & \multicolumn{2}{c}{\textbf{Esophagus}} & \multicolumn{2}{c}{\textbf{Ileum}} \\ \cline{3-9} 
 &  & \textbf{Celiac} & \textbf{EE} & \textbf{Normal} & \textbf{EoE} & \textbf{Normal} & \textbf{Crohn's} & \textbf{Normal} \\ \hline
\multirow{2}{*}{\textbf{Celiac}}&VGGNet & $0.712\mypm 0.083$ &
$0.057\mypm 0.046$ & 
$0.162\mypm 0.058$ &
$0.001\mypm 0.001$ &
$0.001\mypm 0.001$ &
$0.064\mypm 0.025$ &
$0.004\mypm 0.004$ \\ 
 & H-VGGNet & 
 $0.756\mypm 0.051$ &
 $0.026\mypm 0.008$ &
 $0.157\mypm 0.046$ & 
 $0.015\mypm 0.015$ &
 $0.004\mypm 0.004$ &
 $0.032\mypm 0.008$ &
 $0.011\mypm 0.004$ \\ \hline
\multirow{2}{*}{\textbf{EE}} & VGGNet & $0.020\mypm 0.020$ & 
$0.975\mypm 0.025$ &
$0.001\mypm 0.001$ & 
$0.001\mypm 0.001$ &
$0.001\mypm 0.001$ &
$0.004\mypm 0.004$ &
$0.000\mypm 0.000$ \\  
 & H-VGGNet & 
 $0.009\mypm 0.005$ &
 $0.989\mypm 0.005$ &
 $0.001\mypm 0.001$ & 
 $0.000\mypm 0.000$ & 
 $0.001\mypm 0.001$ &
 $0.001\mypm 0.001$ &
 $0.000\mypm 0.000$ \\ \hline
\multirow{2}{*}{\textbf{Normal (Duodenum)}} & VGGNet & 
$0.218\mypm 0.088$ &
$0.035\mypm 0.035$ &
$0.711\mypm 0.071$ &
$0.001\mypm 0.001$ & 
$0.001\mypm 0.001$ &
$0.030\mypm 0.022$ &
$0.007\mypm 0.007$ \\ 
 & H-VGGNet & 
 $0.197\mypm 0.056$ & 
 $0.011\mypm 0.006$ & 
 $0.764\mypm 0.055$ & 
 $0.007\mypm 0.006$ &
 $0.004\mypm 0.003$ & 
 $0.007\mypm 0.002$ &
 $0.011\mypm 0.004$ \\ \hline
\multirow{2}{*}{\textbf{EoE}} & VGGNet & $0.016\mypm 0.006$ & 
$0.004\mypm 0.002$ &
$0.003\mypm 0.003$ &
$0.834\mypm 0.035$ &
$0.123\mypm 0.033$ &
$0.017\mypm 0.008$ &
$0.004\mypm 0.002$ \\ 
 & H-VGGNet & 
 $0.003\mypm 0.001$ & 
 $0.003\mypm 0.001$ &
 $0.001\mypm 0.001$ & 
 $0.920\mypm 0.017$ & 
 $0.066\mypm 0.015$ &
 $0.003\mypm 0.003$ & 
 $0.004\mypm 0.001$ \\ \hline
\multirow{2}{*}{\textbf{Normal (Esophagus)}} & VGGNet & 
$0.009\mypm 0.003$ &
$0.002\mypm 0.002$ &
$0.003\mypm 0.003$ & 
$0.053\mypm 0.013$ & 
$0.915\mypm 0.013$ & 
$0.018\mypm 0.008$ & 
$0.001\mypm 0.001$ \\  
 & H-VGGNet & 
 $0.003\mypm 0.001$ &
 $0.000\mypm 0.000$ & 
 $0.000\mypm 0.000$ & 
 $0.086\mypm 0.022$ & 
 $0.901\mypm 0.023$ &
 $0.007\mypm 0.003$ & 
 $0.003\mypm 0.001$ \\ \hline
\multirow{2}{*}{\textbf{Crohn's}} & VGGNet & 
$0.042\mypm 0.031$ & 
$0.008\mypm 0.008$ & 
$0.017\mypm 0.017$ & 
$0.002\mypm 0.001$ & 
$0.001\mypm 0.001$ & 
$0.928\mypm 0.049$ & 
$0.004\mypm 0.003$ \\  
 & H-VGGNet & 
 $0.019\mypm 0.011$ & 
 $0.002\mypm 0.001$ & 
 $0.009\mypm 0.006$ & 
 $0.007\mypm 0.003$ & 
 $0.003\mypm 0.003$ &
 $0.955\mypm 0.019$ & 
 $0.006\mypm 0.004$ \\ \hline
\multirow{2}{*}{\textbf{Normal (Ileum)}} & VGGNet & 
$0.018\mypm 0.007$ &
$0.019\mypm 0.019$ & 
$0.034\mypm 0.031$ & 
$0.025\mypm 0.018$ & 
$0.031\mypm 0.015$ & 
$0.144\mypm 0.117$ & 
$0.729\mypm 0.093$ \\ 
 & H-VGGNet & 
 $0.011\mypm 0.004$ & 
 $0.004\mypm 0.002$ & 
 $0.007\mypm 0.005$ & 
 $0.057\mypm 0.022$ & 
 $0.025\mypm 0.007$ & 
 $0.144\mypm 0.057$ & 
 $0.752\mypm 0.045$ \\ \hline
\end{tabular}%
}
\end{table*}

\section{Conclusion}\label{sec:Conclusion}
This paper proposes a hierarchical deep convolutional neural network for multi-category classification of GI disorders using histopathological biopsy images. Our proposed model was tested on~$25,582$ cropped images derived from an independent set of $373$ WSIs. Our results showed that the hierarchical model had superior classification performance for a problem with an inherent hierarchical structure compared to a flat model, which assumes equal difficulty for classification. With the dataset collected from $n=441$ patients and based on our training, development, and test set split, our model can be generalized to other patients that are not part of the training or development sets. 

In CNN architecture, since lower layers capture low-level features while higher layers are likely to extract more abstract features, we utilized this property to build our model instead of employing separate models for different class hierarchy levels. The use of such a structure makes it possible to save computational cost and benefit from shared information across the coarse levels in the training phase. Quantification of this synergy could be a possible avenue for future research.

\vspace{10pt}

\section*{Acknowledgements}
The research reported in this manuscript was supported by the National Institute of Diabetes and Digestive and Kidney Diseases of the National Institutes of Health under award number K23DK117061-01A1 (SS), Bill and Melinda Gates Foundation under award numbers OPP1066203, OPP1066118, OPP1144149 and OPP1066153 and University of Virginia Translational Health Research Institute of Virginia~(THRIV) Scholar Career Development Award (SS). The content is solely the authors’ responsibility and does not necessarily represent the official views of the funding agencies. 

\bibliographystyle{IEEEtran} 
\bibliography{refs}

\end{document}